\documentclass[eclepsf,showpacs,preprintnumbers,amsmath,amssymb]{revtex4}

\newcommand{\simgt}{\lower.5ex\hbox{$\; \buildrel > \over \sim \;$}}
\newcommand{\simlt}{\lower.5ex\hbox{$\; \buildrel < \over \sim \;$}}

\usepackage{graphicx}
\usepackage{dcolumn}
\usepackage{bm}

\begin{document}


\title{Wave effect in gravitational lensing by a cosmic string
}

\author{Kazuhiro Yamamoto and Keiji Tsunoda}
\affiliation{
Department of Physical Science, Hiroshima University,
Higashi-Hiroshima 739-8526,~Japan}



\begin{abstract}
The wave effect in the gravitational lensing phenomenon by a straight 
cosmic string is investigated. The interference pattern is expressed 
in terms of a simple formula. We demonstrate that modulations of the
interfered wave amplitude can be a unique signature of the wave 
effect. We briefly mention a possible chance of detecting the wave 
effect in future gravitational wave observatories. 
\end{abstract}

\maketitle


\def\M{{M}}
\section{Introduction}

The standard model of elementary particle theory involves
the mechanism of spontaneous symmetry breaking, which 
predicts phase transitions in the early universe. 
In general, such phase transitions produce topological
defects, depending on the symmetry spontaneously broken. 
The cosmic string is the topological defect produced when 
a $U(1)$ symmetry is broken \cite{KT}. 
Historically, the cosmic 
strings were well investigated, motivated by a possible
mechanism to explain the structure formation of the universe 
\cite{Stringformation}.
However, this possibility is severely limited or rejected  
by the observation of large scale structure of the 
galaxies and the measurements of the cosmic microwave 
background (CMB) anisotropies \cite{WMAP,2dF}.
Thus the cosmic strings cannot play a central role in
structure formation, which constrains the cosmic string 
energy density. But, we also note that a cosmological model with a 
non-negligible defect contribution has not been rejected completely.
It is argued that the model with a mixture of inflation 
and topological defects can be compatible with a CMB anisotropy 
data \cite{BPRS}.
 
On the other hand, other possible roles of the cosmic strings are 
considered. For example, Sazhin et al. have reported 
detection of the source which is naturally explained 
by lensing effect of a cosmic string \cite{Sazhin}. 
The gravitational lensing effect due to the cosmic strings
has been investigated by several authors
\cite{Vilenkin,Gott,LV,DEL,BU}. Actually the gravitational 
lensing phenomenon is known as a possible unique probe to detect 
the cosmic strings. The prospect to detect the cosmic string with 
wide field galaxy surveys is discussed \cite{HV,Shirasaki}.
In the present paper, we investigate the wave effect in the 
gravitational lensing by the cosmic string. 
Recently wave effect in gravitational lensing
has been investigated, motivated by the gravitational
wave observation \cite{TN,Seto,PINQ,BHN,TTN}. 
These works focus on wave effect by isolated lenses
(see also \cite{SEF,Deguchi}). However, the wave effect
in the lensing by the cosmic string has not been investigated,
as far as we know. 

This paper is organized as follows:
In Sec. 2 we present a solution of the wave equation
on the background spacetime with a straight static cosmic string.
Then we show the condition that the interference occurs as a
wave effect. 
In Sec. 3, we demonstrate typical numbers of gravitational 
lensing by the cosmic string.
We also discuss a possibility of detecting the interference in the 
lensing of gravitational waves from a compact binary in Sec. 4. 
Sec. 5 is devoted to conclusions.
We use the convention $G=c=1$, otherwise we express them explicitly. 
\section{Wave Solution}
\def\dls{{D_{\rm LS}}}
\def\dos{{D_{\rm OS}}}

The properties of a cosmic string in a vacuum spacetime
was studied by Vilenkin \cite{VilenkinA}. He showed that
the local nature of the solution is same as that of the 
Minkowski spacetime, but the conical singularity arises 
around the cosmic string. 
This is the origin of the deficit angle 
$\delta=8\pi \mu$ of the spacetime, where $\mu$ is the
energy (line) density of the cosmic string.
The energy density $\mu$ depends on the individual model of the 
cosmic string, however, the constraint on $\mu$
from cosmological observations of the galaxy large scale
structure and the cosmic microwave background anisotropies
is $\mu\simlt 10^{-6}$ (\cite{HV} and references therein).

We consider the simple configuration (see Fig. 1):
A straight string is located parallel to the $z$ axis.
The source is located on the coordinate $A=(-L\cos[\delta/2],L\sin[\delta/2],0)$
and $B=(-L\cos[\delta/2],-L\sin[\delta/2],0)$, where both $A$ and $B$ are identical
because this spacetime has the conical singularity.
Denoting the coordinate of an observer $(x,y,z)$,
the distance between $A (B)$ and the observer is 
\begin{eqnarray}
  &&r_1=\sqrt{\left(x+L\cos{\delta\over 2}\right)^2
       +\left(y-L\sin{\delta\over 2}\right)^2+z^2},
\label{ra}
\\
  &&r_2=\sqrt{\left(x+L\cos{\delta\over 2}\right)^2
       +\left(y+L\sin{\delta\over 2}\right)^2+z^2},
\label{rb}
\end{eqnarray}
respectively. We assume that the amplitude of the wave field is expressed as
\begin{eqnarray}
  {\cal E}
  &=&A_0{\cos(\omega t-kr_1)\over r_1}\Theta(y+x\tan{\delta/2})  
          +A_0{\cos(\omega t-kr_2)\over r_2}\Theta(y-x\tan{\delta/2}),
\label{calEE}
\end{eqnarray}
where $\Theta(X)$ is the Heaviside function and $A_0$ is a constant.  
The interference occurs in the region
$y\ge -x\tan{\delta/2}$ and $y\le x\tan{\delta/2}$. 
Within this region we have  
\begin{eqnarray}
  {\cal E}
   &\simeq&{2A_0\over r}\cos\left[\left(\omega t-k{r_1+r_2\over2}\right)\right]
      \cos\left[{k(r_1-r_2)\over2}\right]
\end{eqnarray}
with $r=\sqrt{(x+L)^2+z^2}$. Thus the amplitude of the wave 
is determined by the phase $k(r_1-r_2)/2$ because of the interference.
Using eqs. (\ref{ra}) and (\ref{rb}), the path difference is 
\begin{eqnarray}
 r_1-r_2&\simeq& {2y\sin[\delta/2]} {L\over \sqrt{(x+L)^2+z^2}}
={2y \sin[\delta/2]}{ \dls \sin\theta \over \dos},
\label{rr}
\end{eqnarray}
where we assumed $y\ll z,x,L$, in the last equality, we used
$\theta$ to denote the angle between  the string direction 
and the line of sight, and $\dls$ and $\dos$ to denote 
angular diameter distances from observer to source and 
string to source, respectively.
The condition that the wave amplitude has a maximum is
\begin{eqnarray} 
k(r_1-r_2)/2=n\pi
\label{krr}
\end{eqnarray} 
with integer $n=0,\pm1,\pm2,\cdots \pm n_{\rm max}$,
where $n_{\rm max}$ is the largest integer satisfying 
\begin{eqnarray}
\pi|n|\le kx \sin^2[\delta/2] { \dls \sin\theta \over \dos}.
\end{eqnarray}

\section{Typical numbers}
\def\Hz{{\rm Hz}}

In this section we estimate quantitatively physical 
numbers of the gravitational lensing system by the cosmic 
string. It is well known that the angular separation of 
two images is
\begin{eqnarray}
  \Delta \alpha= 2\sin[\delta/2] {\dls\sin\theta\over\dos}
  \simeq 2''
  \left({\delta\over 10^{-5}}\right)
  \left({\dls\sin\theta\over\dos}\right).
\end{eqnarray}
The path difference causes time delay, defined by 
$T_{dl}={|r_1-r_2|/c}$. To estimate its typical number, 
we set $y=(1+z_L)D_{\rm LO}\sin[\delta/2]/2$ as a typical
value in eq. (\ref{rr}), where $D_{\rm LO}$ is the angular
diameter distance between the string and observer. Then, 
we have
\begin{eqnarray}
  T_{dl}&=&{(1+z_L)\sin^2[\delta/2]D_{\rm LO}\over c} \left({\dls\sin\theta\over \dos}\right)
\nonumber
\\
  &\simeq&10^7 (1+z_L) \left({\delta \over 10^{-5}}\right)^2
  \left({D_{\rm LO} \over cH_0^{-1}}\right)
  \left({\dls\sin\theta\over\dos}\right)~{\rm sec}
\end{eqnarray}
with the Hubble constant $H_0=70~{\rm km/s/Mpc}$. 

On the other hand, eq. (\ref{krr}) is the condition for 
the maximum peak of the wave. Therefore the wave effect is 
characterized by the length between two neighboring peaks,
\begin{eqnarray}
  \Delta y={\pi\over k\sin[\delta/2]}
  \left({\dos\over\dls\sin\theta}\right) \simeq 3\times 10^{15}
  \left({1{\Hz}\over \nu}\right)
  \left({10^{-5}\over \delta}\right)
  \left({\dos\over\dls\sin\theta}\right)~{\rm cm},
\end{eqnarray}
where $\nu$ is the frequency of the wave. Thus $\Delta y$ can 
be an astronomical distance and the observed amplitude of the
wave changes in this periodic interval due to the wave effect. 
If the string moves in a direction perpendicular to the line
of sight direction, the amplitude of the wave will modulate
periodically. Assuming
that the cosmic string moves in the direction of $y$ axis
with a speed near the velocity of light, 
the typical period of the modulation is roughly estimated as 
\begin{eqnarray}
  {\cal T}_v={\Delta y\over c} \simeq  10^7 
  \left({0.01{\Hz}\over \nu}\right)
  \left({10^{-5}\over \delta}\right)
  \left({\dos\over\dls\sin\theta}\right)~{\rm sec}.
\end{eqnarray}

The statistical probability of the gravitational lensing
by a cosmic string can be estimated as follows:
The probability for lensing for a single source   
at the redshift $z_S$ due to infinitely long cosmic 
string located at the redshift $z_L$ is \cite{HV}
\begin{eqnarray}
  P(z_S,z_L)\sim {2\delta\over \pi^2}{\dls\over\dos}
  \simeq 2\times10^{-6} \left({\delta\over 10^{-5}}\right)
  \left({\dls\over\dos}\right).
\end{eqnarray}
Thus the probability for a cosmic string is very small. 
However, numerical simulations of string networks show the
possibility that more strings can exist within the horizon 
effectively \cite{DEL,BB}.

\section{Wave effect in gravitational waves}
\def\calT{{\cal T}}

Now let us consider a gravitational wave from a massive binary 
with an equal mass $M$ as a source. We assume that the
binary angular frequency $\bar \omega$ corresponds to the
frequency of the gravitational wave $\nu$ of an observer 
by $\bar\omega=2\pi\nu(1+z_S)$. 
The lifetime of the binary system $T_{lf}$ must be longer 
than the time delay $T_{dl}$ so that the interference occurs. 
The binary lifetime is estimated \cite{Schutz}
\begin{eqnarray}
  T_{lf}\sim 0.02M^{-5/3} \bar\omega^{-8/3}\simeq{1.3\times 10^5\over (1+z_S)^{8/3} }
  \left({M\over M_{\odot}}\right)^{-5/3}
  \biggl({\nu \over {\rm Hz}}\biggr)^{-8/3}~{\rm sec} .
\end{eqnarray}
Then, the condition $T_{lf}>T_{dl}$ yields
\begin{eqnarray}
  {M}&<&{120\over (1+z_L)^{3/5}(1+z_S)^{8/5}}
\nonumber
\\
  &&\times\biggl({\nu\over 0.01 \Hz}\biggr)^{-8/5}
  \left({\delta\over 10^{-5}}\right)^{-6/5}
  \left({D_{\rm LO}\over c H_0^{-1}}\right)^{-3/5}
  \left({\dls \sin\theta \over \dos}\right)^{-3/5}~ M_{\odot}.
\label{Mmax}
\end{eqnarray}
The DECIGO planed as a future gravitational wave observatory has an 
excellent sensitivity around $\nu=0.1~\Hz$ \cite{SKN}. 
For $\nu=0.01~{\rm Hz}$ and $\delta=10^{-5}$, eq. (\ref{Mmax}) 
yields the maximum mass of the binary larger than $ 1 M_{\odot}$. 
Thus the neutron stars (NS-NS) binary can be the source for the 
wave effect.

Due to emission of the gravitational waves, the binary orbit
changes, then the angular frequency of the wave changes. 
Namely, waves with different angular frequencies
interfere because of the time delay effect. This causes 
another modulation of the wave amplitude as a wave effect, 
whose period is approximately written as 
\begin{eqnarray}
  \calT_{p}&=&\displaystyle{{4\pi \over (r_1-r_2){d\omega/dt}}}
\nonumber
\\  
  &\simeq&{1.5\times 10^{6}\over (1+z_L)(1+z_S)^{5/3}}
  \biggl({\nu\over 0.01\Hz}\biggr)^{-11/3}
  \left({M\over M_{\odot}}\right)^{-5/3}
  \left({ 10^{-5}\over \delta}\right)^{2}
  \left({c H_0^{-1}\over D_{\rm LO}}\right)
  \left({\dos\over \dls \sin\theta }\right)~{\rm sec}.
\nonumber
\\  
\end{eqnarray}
For $\nu\simgt 0.01~\Hz$ and $M\sim M_\odot$, ${\cal T}_{p}$ is 
shorter than the period of the modulation due to the string motion ${\cal T}_{v}$.

\section{Conclusion}
In this work we have investigated wave effect in gravitational
lensing by a cosmic string. Our investigation is limited to the simplest
case of the static and straight cosmic string. In this case the 
interference of the wave field is expressed in a very simple form eq. 
(\ref{rr}). The condition that the interference occurs is eq.
(\ref{krr}). When the string moves in the direction perpendicular 
to the line of sight direction, the wave amplitude for an observer
will modulate by a wave effect in gravitational lensing.
A chirp gravitational wave from a compact binary at a cosmological 
distance is a possible source in which we might detect the wave effect.
However, concerning this source, the change of orbit due to  
gravitational wave emission causes another modulation of the wave amplitude. 
These modulations can be a unique signature of the cosmic string 
when a detector with sufficient angular resolution and sensitivity 
is assumed. 
Due to the same mechanism, a similar modulation can appear in a chirp 
gravitational wave by the gravitational lensing by a galaxy halo \cite{Yamamoto}.  
However, in this case, the flux ratio of two waves must be almost  
$1$ for the clear modulation. Namely, a configuration of a lensing system 
near the Einstein ring must be required. This means that the lensing 
probability is relatively small and the flux is amplified by the
magnification effect. Thus, in principle, it can be discriminated 
whether the lens causing the modulating signal is a cosmic string or 
a galaxy halo.

In the present work, we have considered an idealized situation.
We have assumed a point source, neglecting the finite size effect 
of the source. For the gravitational wave from a compact binary, 
however, the assumption will be plausible. We have considered
the interference of two waves with the same amplitude. The change 
of the orbit of compact binary results in a difference between the
two wave amplitudes due to the time delay effect. However, it can
be negligible for a low frequency binary $\nu\sim 0.01~\Hz$
with $M\sim M_\odot$. 

Various sources of gravitational waves and their detectability 
have been considered (e.g., \cite{Hughes} for a recent review). 
Binaries of massive objects will be the most stable source of a 
periodic wave, and NS-NS binary is the most promising source.
The ground based detectors, such as LIGO and VIRGO, have the sensitivity
in the high frequency band, $1~$Hz$\simlt f\simlt 10^4~$Hz, which
corresponds to that of gravitational waves from the NS-NS binary
at final coalescing stage. 
On the other hand, the gravitational wave observatories in space, 
such as LISA and DECIGO, will detect gravitational waves in the low 
frequency band, $10^{-4}~$Hz$\simlt f\simlt 1~$Hz.
The modulation  will not be observed by the facility with the high 
frequency band because it appears in stable and periodic gravitational 
waves. However, the facility with the low frequency band might have a 
chance to observe the modulation, if the sensitivity was sufficient. 
The detectability of a signal depends on the distance of a 
source and a method of data analysis, therefore more details 
are needed for the definite conclusion. Actually, when the angular
resolution is not sufficient, modulations of gravitational waves 
can appear due to superposition of waves from different individual 
binary sources.

\vspace{1mm}
{\it Acknowledgments} 
The authors thank Y. Kojima for reading the manuscript and 
useful discussions and comments. K.Y. thanks T. Hayashino 
for useful communication.
This work is supported in part by Grant-in-Aid for
Scientific research of Japanese Ministry of Education, 
Culture, Sports, Science and Technology, Grant No.15740155.


\newpage
\begin{figure*}
\includegraphics{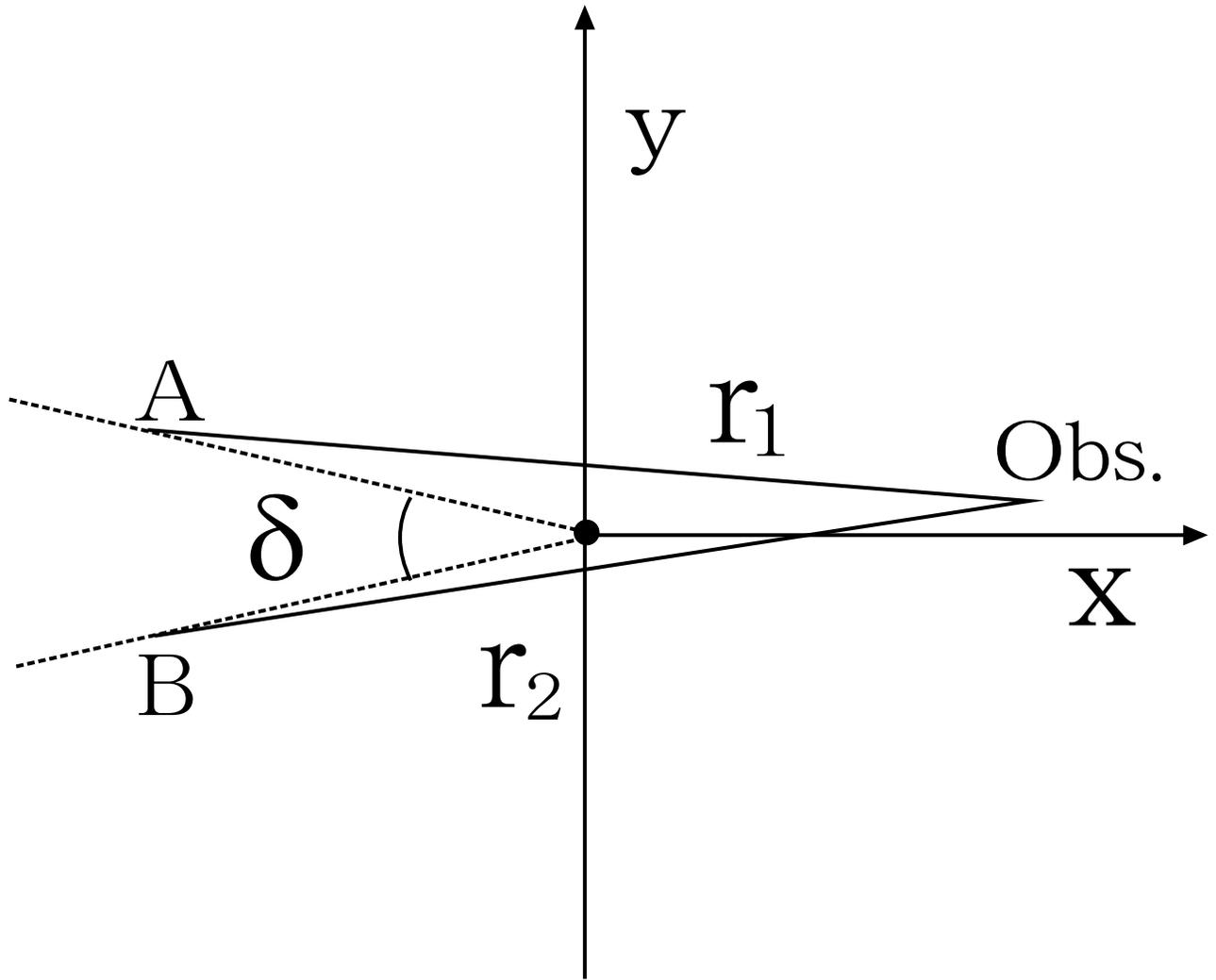}
\caption{\label{fig1} Configuration of the lensing system:
The cosmic string is located parallel to the $z$ axis.
This figure shows a projection on the $x-y$ plane. 
The dashed two lines should be identified due to
conical singularity with the deficit angle $\delta$.}
\end{figure*}

\end{document}